# Adaptive Extensive Cancellation Algorithm and Harmonic Enhanced Heart Rate Estimation based on MMWave Radar

Hui Tang*, *Member, IEEE*, Zhan Yang, *Student*, Yu Rong, *Member, IEEE*, and Li Chai, *Member, IEEE*

*Abstract* —Heart rate (HR) monitoring is crucial for assessing physical fitness, cardiovascular health, and stress management. Millimeter-wave radar offers a promising non-contact solution for long-term monitoring. However, accurate HR estimation remains challenging in low signal-to-noise ratio (SNR) conditions. To deal with both respiration harmonics and intermodulation interference, this paper proposes a cancellation-before-estimation strategy. Firstly, we present the adaptive extensive cancellation algorithm (ECA) to suppress respiratory and its low-order harmonics. Then, we propose an adaptive harmonic enhanced trace (AHET) method to avoid intermodulation interference by refining the HR search region. Various experimental results validate the effectiveness of the proposed methods, demonstrating improvements in accuracy, robustness, and computational efficiency compared to conventional approaches based on the FMCW (Frequency Modulated Continuous Wave) system.

## I. INTRODUCTION

HEART rate (HR) monitoring plays a crucial role in assessing cardiovascular health, tracking fitness levels, managing stress, and evaluating post-exercise recovery [1]–[4]. It can detect underlying health problems, guide the intensity of training, and assist individuals with medical conditions in managing their treatment effectively. Currently, clinical HR monitoring relies primarily on contact-based technologies such as electrocardiograms (ECG), pulse wave detectors, and HR monitors [5], [6]. Although these methods provide accurate cardiac data, they require wearable devices that may cause discomfort or inconvenience to patients. In addition, concerns about privacy and skin irritation are particularly relevant for newborns and individuals with skin conditions. Non-contact technologies offer a promising alternative, addressing these limitations and representing an important direction for future cardiac monitoring. Compared to other non-contact methods, such as cameras, WiFi routers, and acoustic sensors, radar-based systems have several advantages: they are less affected by changes in lighting, temperature, and sound, and they do not compromise privacy. These benefits make radar a highly promising technology for intelligent HR monitoring [7], [8]. With the advent of commercial radar platforms, especially those operating at 66 GHz and 77 GHz, millimeter-wave radars based on the FMCW (Frequency Modulated Continuous Wave) system have become the mainstream solution for HR monitoring. Several universities and research institutions have developed experimental systems and published open datasets to advance this technology [9], [10].

Although radar-based HR monitoring offers significant advantages, continuous monitoring of fundamental heartbeat (HB) faces critical challenges, primarily due to spurious peaks in the spectrum such as high-order harmonics caused by respiration [11]–[13]. HB signals typically exhibit small displacements of the chest wall (0.1 to 0.5 millimeters) and a frequency range of 0.7 to 2 Hz, while respiratory signals involve larger displacements (1 to 5 mm) and lower frequencies (0.1 to 0.5 Hz) [14]. The overlap between respiratory harmonics and the fundamental HB spectrum often leads to failed HR estimation, as HB peaks can be masked by respiratory harmonics. This harmonic interference has been widely recognized as a major challenge in HR monitoring [8], [15]. Furthermore, existing methods struggle to address additional interferences caused by radar nonlinearity or phase-demodulation issues [16]–[18].

To circumvent these challenges, researchers have explored harmonic-based approaches. Studies [19] [20] suggest that while the fundamental HB signal is limited by respiratory interference, its high-order harmonics are constrained by noise. Accurate HR estimation can be achieved by identifying the second-order HB harmonic within the harmonic search region. In [16], an adaptive NLS algorithm combined with a Kalman filter was proposed to select the best HR estimate from both fundamental and harmonic search regions. However, under low signal-to-noise ratio (SNR) conditions, these methods require further investigation, especially when both the fundamental

Manuscript received Month xx, 2xxx; revised Month xx, xxxx; accepted Month x, xxxx. This work was supported by the National Natural Science Foundation of China under grants 62173259 and 62473298 and Zhejiang Province Natural Science Foundation LZ24F030006.

Hui Tang is the corresponding author; Hui Tang and Zhan Yang are with the Engineering Research Center of Metallurgical Automation and Measurement Technology, Wuhan University of Science and Technology, Wuhan, 430081, China (e-mail: htang@wust.edu.cn, YangZ@wust.edu.cn).

Yu Rong is with Center for Wireless Information Systems and Computational Architectures (WISCA) at the School of Electrical, Computer and Energy Engineering, Arizona State University, Tempe, AZ 85281 USA, e-mail: yrong5@asu.edu.

Li Chai is with the College of Control Science and Engineering, Zhejiang University, Hangzhou, 310027, China, email: chaili@zju.edu.cn

and harmonic HB are simultaneously affected by respiratory harmonics and intermodulation components. Intermodulation interference (defined here as the sum or difference of the fundamental HB and respiratory signals or their harmonics) remains an underexplored issue in existing literature. Reported methods need reevaluation under conditions where intermodulation interference significantly impacts HR estimation.

To address this issue, we propose a cancellation-before-detection strategy to improve HR estimation accuracy and robustness. First, we introduce the adaptive Extensive Cancellation Algorithm (ECA) to suppress respiration and its harmonics, thereby enhancing the detection of the fundamental HR peak. Following interference cancellation, an Adaptive Harmonic Enhanced Trace (AHET) method is developed to mitigate intermodulation interference and further improve HR detection accuracy. Overall, this key contributions of this work are as follows,

1) Propose an adaptive ECA algorithm to eliminate respiration and its harmonics. The adaptive NLS (ANLS) algorithm is utilized to reconstruct the reference respiration.
2) Develop an AHET method to mitigate intermodulation interference to enhance the HR detection by leveraging the harmonic HB. Credibility assessment is defined to design a fine and reliable HR search region.
3) Conduct extensive real-world experiments to verify the effectiveness and advantage of the proposed method.

The organization of the rest of this article is as follows: The task of the problem studied in this article is formulated in Section II. Section III introduces the signal model. The proposed method is well described in Section IV including the adaptive ECA algorithm and the AHET method. Then, various experiments are designed in Section V to verify the effectiveness of the proposed method. Finally, Section VI concludes this paper.

## II. Problem Formulation

The primary objective of this article is to enhance the probability of accurately detecting HR peaks and to improve the robustness of HR estimation. This is achieved through two key techniques: (1) eliminating respiration and its harmonics, and (2) designing a refined search region to avoid intermodulation interference. Intermodulation interference and high-order respiratory harmonics can be closely spaced or even merge, forming strong competing peaks near HR, like $f_1 \approx 1^{st}HR - 1^{st}RR$, $f_2 \approx 1^{st}HR + 1^{st}RR$, $f_3 \approx 1^{st}HR + 2 \cdot 1^{st}RR$. If the frequency resolution is insufficient to distinguish between intermodulation interference and respiratory harmonics, these peaks may merge into a stronger one, masking the HR peak and causing detection failures.

Fig.1 illustrates the masking effects of respiratory harmonics and intermodulation interference on both the fundamental and second-order harmonic regions using measured data. Fig.1(a-1) and (b-1) demonstrate the masking effect within the fundamental search region (42 to 120 BPM) for reference HRs of 76.6 BPM and 70.6 BPM, respectively. In both cases, it is necessary to eliminate respiration and up to the third-order harmonics. In Fig.1(a-1), the RR and HR are 15.6 BPM and 76.6 BPM, respectively. The intermodulation interference $f_1$ is 61 BPM and the forth-order RR is 62.4 BPM. Their 1.4 BPM difference, being smaller than the 3 BPM frequency resolution, causes the peaks to merge, forming a stronger additional peak at 62.7 BPM (marked by the blue arrow). This peak competes with the fundamental HR peak at 76.6 BPM (marked by the red arrow). Fig.1(b-1) presents another case where RR and HR are 15.6 BPM and 70.6 BPM, respectively. The intermodulation interference $f_1$ is 55 BPM and the forth-order RR is 62.4 BPM. Their 7.4 BPM difference exceeds the frequency resolution, resulting in two separate peaks (marked by blue arrows), both weaker than the fundamental HR peak at 70.6 BPM. Fig.1(a-2) and (b-2) show the masking effect on harmonic search region (from 84 BPM to 180 BPM). In Fig.1(a-2), the intermodulation interference $f_2$ and $f_3$ occur at 92.2 BPM and 107.8 BPM, respectively. These peaks merge with $6^{th}$-order RR (93.6 BPM) and $7^{th}$-order RR (109.2 BPM) due to their proximity, creating strong competing peaks (marked by blue arrows) that hinder harmonic HR detection. In Fig.1(b-2), the intermodulation interference $f_2$ and $f_3$ are 86.2 BPM and 101.8 BPM, which may be separated with $6^{th}$-order RR (93.6 BPM) and $7^{th}$-order RR (109.2 BPM), respectively. These peaks are separated and visible in (b-2). Although $f_2$ or $f_3$ are still comparable to the harmonic HR peak, the clear separation allows for more reliable detection. Both (b-1) and (b-2) demonstrate that a refined search region is effective in mitigating intermodulation interference and enhancing HR detection.

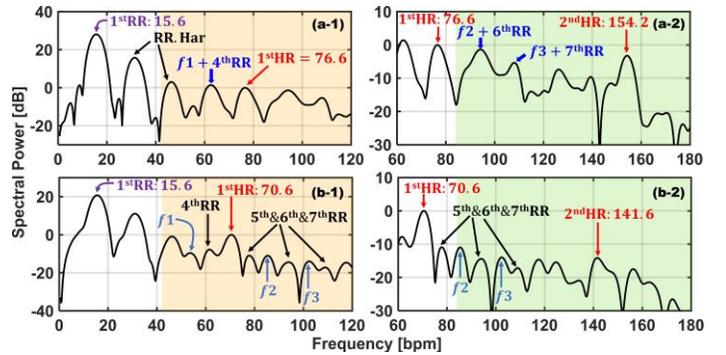

Fig. 1. The masking effect of respiratory harmonics and intermodulation interference (i.e. $f_1, f_2, f_3$), in which $f_1 \stackrel{=}{=} 1^{st}HR - 1^{st}RR$, $f_2 \approx 1^{st}HR + 1^{st}RR$, $f_3 \stackrel{=}{=} 1^{st}HR + 2 \cdot 1^{st}RR$. The orange and green patches represent the search ranges for $1^{st}HR$ and $2^{nd}HR$, respectively. (a-1) and (a-2) demonstrate $1^{st}HR$ (76.6 BPM) is obstructed by $3^{rd}RR$ and $f_1 + 4^{th}RR$ and $2^{nd}HR$ is obstructed by $f_2 + 6^{th}RR$ and $f_3 + 7^{th}RR$, due to the limitation of frequency resolution (20s CPI corresponds to a frequency resolution of 3 bpm). (b-1) and (b-2) demonstrate $1^{st}HR$ (70.6 BPM) is obstructed by $3^{rd}RR$ and $2^{nd}HR$ is obstructed by intermodulation interference $f_2$ and $f_3$.

In general, it is necessary to cancel respiration and up to third-order harmonics to improve the detection performance in fundamental region under any circumstances. To avoid intermodulation interference and high-order RR harmonics, it is efficient to design a refined and well-defined search region. This paper thoroughly investigates these two problems.

## III. System Model

The transmitted FMCW can be written as

$$x_T(t) = \sqrt{A_T}\cos(2\pi f_c t + \pi \frac{B}{T_c} t^2 + \phi(t)), \quad (1)$$

where $A_t$ is the transmitted power, $f_c$ is the operating frequency, $B$ is the pulse bandwidth, $T_c$ is the "chirp" duration, and $\phi(t)$ is the transmitter's phase noise of local oscillator.

The chest wall displacement signal of a subject at a nominal distance $d_o$ can be modeled as a sum of two periodic signals from respiratory and cardiac activities and their harmonics,

$$d(t) = d_o + \sum_{k=1}^{K_b} \alpha_k \sin(2\pi k f_b t + \psi_k) + \sum_{l=1}^{K_h} \beta_l \sin(2\pi l f_h t + \delta_l), \quad (2)$$

where $K_b$ and $K_h$ are the harmonically related sinusoids for breathing and heartbeat signal. $f_b$ and $f_h$ are fundamental frequencies related to breathing and heartbeat. $\alpha_k$ and $\beta_l$ are the amplitudes of the $k$th respiratory harmonic and the $l$th heartbeat harmonic. $\psi_k$ and $\delta_l$ are the initial phases of the $k$th respiratory harmonic and the $l$th heartbeat harmonic, respectively.

The received signal can be modeled as a sum of the target response and the delayed, attenuated versions of the transmitted pulse due to static environment,

$$x_R(t) = M_T x_T(t - t_D) + \sum M_i x_T(t - t_i), \quad (3)$$

where $M_T$ and $M_i$ denote the magnitudes of the target response and the multi-path components, while $t_D = 2d(t)/c$ and $t_i$ are the corresponding delays, where $c$ is the speed of light.

The signal of interest can be modeled as,

$$x_o(t) = M_T x_T(t - t_D) \quad (4)$$
$$= M_T \sqrt{A_T}\cos(2\pi f_c(t - t_D) + \pi \frac{B}{T_c}(t - t_D)^2 + \phi(t - t_D)),$$

By mixing the received signal with the transmitted signal, the intermediate frequency (IF) signal $x_{IF}(t)$ is obtained after low-pass filtering. In short-distance measurements, the IF signal can be simplified as follows,

$$x_{IF}(t) = A_{IF}\exp(j(2\pi f_{IF}t + \triangle\theta)). \quad (5)$$

where $f_{IF}$ is the frequency of the IF signal, $f_{IF} = Bt_D/T_c$. $\triangle\theta = 2\pi f_c t_D$ denotes the time-varying phase relative to the movement of the monitored subject.

After an analog-to-digital converter (ADC), the time-varying phase associated with an object, influenced by displacements from breathing and heartbeat, can be modeled as,

$$\theta[n] = \frac{4\pi}{\lambda_c}(d_b[n] + d_h[n]), \quad (6)$$

$d_b[n]$ and $d_h[n]$ are the displacements due to breathing and heartbeat, respectively. $\lambda_c = c/f_c$ is the operating wavelength.

Let us consider the phase relative to breathing and define $\mathbf{d}_b = [d_b[0],\ldots,d_b[N-1]]^T \in R^N$, the vector consisting of $N$ consecutive samples of $d_b[n]$, which can be expressed as

$$\mathbf{d}_b = \mathbf{H}_b \boldsymbol{\alpha}_b, \quad (7)$$

with $\boldsymbol{\alpha}_b \in R^{K_b}$ being the vector containing the amplitudes of the breathing harmonics and the matrix $\mathbf{H}_b \in R^{N \times K_b}$ having a Vandermonde structure, being constructed from the $K_b$ sinusoidal vectors as

$$\mathbf{H}_b = \begin{bmatrix} \sin(2\pi f_b + \psi_1) & \cdots & \sin(2\pi \cdot K_b f_b + \psi_{K_b}) \\ \vdots & \ddots & \vdots \\ \sin(2\pi f_b(N-1) + \psi_1) & \cdots & \sin(2\pi \cdot K_b f_b(N-1) + \psi_{K_b}) \end{bmatrix} \quad (8)$$

Using this definitions, the phase model in (6) can be rewritten as

$$\boldsymbol{\theta} = \frac{4\pi}{\lambda_c}(\mathbf{H}_b \boldsymbol{\alpha}_b + \mathbf{H}_h \boldsymbol{\beta}_h), \quad (9)$$

with $\boldsymbol{\beta}_h \in R^{K_h}$ being the vector containing the amplitudes of the breathing harmonics and the matrix $\mathbf{H}_h \in R^{N \times K_h}$.

## IV. PROPOSED METHOD

This section outlines the main steps of the proposed method. Fig. 2 presents the overall process framework for HR monitoring using FMCW radar. This framework consists of three key components: (1) data preprocessing for extracting vital signs from the FMCW radar echo signal, (2) an adaptive ECA to suppress low-order respiratory harmonic interference, and (3) an AHET method to deal with intermodulation interference. This framework improves the robustness and accuracy of HR estimation, effectively mitigating challenges such as respiratory harmonics and intermodulation interference, thereby making it suitable for continuous HR monitoring in low SNR scenarios.

### A. Data Preprocessing

Using the fast Fourier transform (FFT) to locate the target and applying the average cancellation method to remove static clutter, a range-time map is generated to identify the target range bin. The phase information of chest wall movement is then extracted using arctangent demodulation [18] and phase unwrapping techniques. The resulting displacement signal can be expressed as follows:

$$d(n) = \frac{\lambda_c}{4\pi} \cdot \text{unwrap}\left[\arctan\left(\frac{Q(n)}{I(n)}\right)\right], \quad (10)$$

where $I(n)$ and $Q(n)$ represent the real and imaginary samples of the complex slow-time signal at the target range bin. However, the FMCW radar achieves high range resolution, enabling the human target to be treated as an extended target spanning multiple range bins [21] [22].

To enhance the information on target chest displacement extracted from the phase signal, the slow-time phase correlation method proposed in [16] is applied to improve the SNR before estimation.

### B. Adaptive ECA Algorithm

To eliminate the respiration and its low-order harmonics, adaptive ECA algorithm is proposed. The ECA is a widely used method for clutter cancellation in passive radar [23]. In traditional ECA algorithms, the reference interference is typically assumed to be known. However, in this paper, the desired respiration interference is unknown, making the construction of respiration and its harmonics a critical challenge to solve. The ANLS method is employed to reconstruct respiration and its harmonics [24].

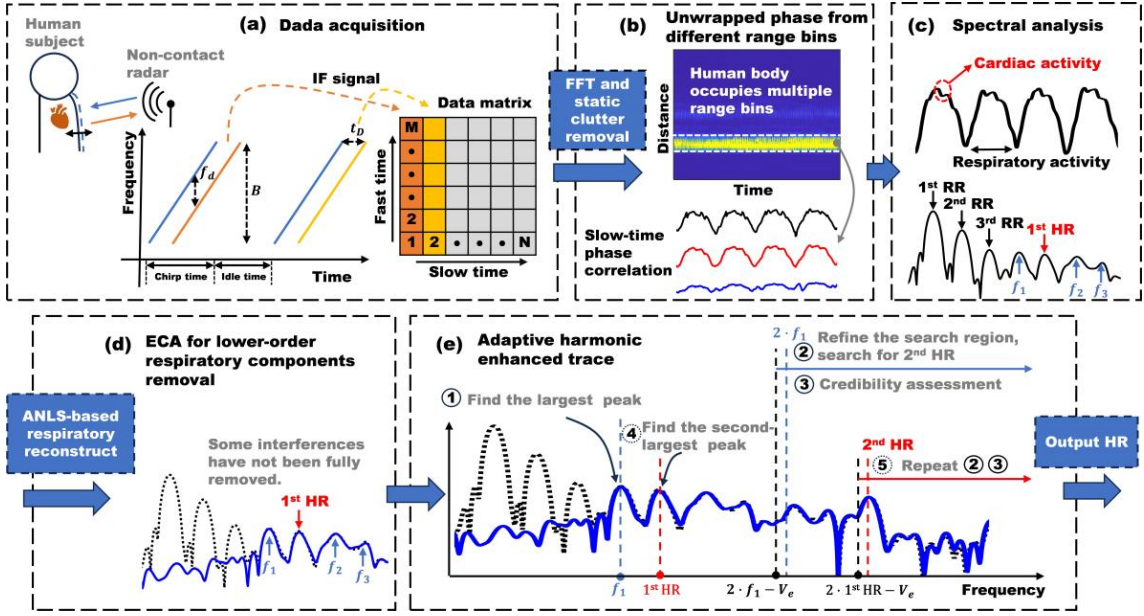

Fig. 2. Overview of the proposed cancellation before estimation HR monitoring framework: (a) and (b) is data preprocessing for extracting vital signs; (c) is spectral analysis; (d) is the proposed adaptive ECA method and (e) is the proposed AHET algorithm.

The signal model is formulated by minimizing the residual signal power after cancelling the disturbance (respiration and its harmonics),

$$\min_{\mathbf{w}} \|\theta - \mathbf{Xw}\|^2, \quad (11)$$

where $\mathbf{X}$ is the subspace matrix of respiration and its $k_b$ harmonics

$$\mathbf{X} = \begin{bmatrix} s_{ref}(0) & \cdots & s_{ref}(N-M+1) \\ \vdots & \ddots & \vdots \\ s_{ref}(N-1) & \cdots & s_{ref}(N-M+1) \end{bmatrix}, \quad (12)$$

where $M$ is the order of the filter, and $\mathbf{s}_{ref} = \mathbf{H}_b \boldsymbol{\alpha}_b \in \mathbb{R}^N$, which is unknown in the considered situation.

To obtain the NLS estimates of respiration, we search for the set of fundamental frequencies that minimize the difference between the recovered displacement signal $\hat{s}_{ref}$ and the signal model in (7). Thus, the NLS optimization problem can be formulated as

$$\min_{\boldsymbol{\alpha}_b, f_b} \|\hat{\mathbf{s}}_{ref} - \mathbf{H}_b \boldsymbol{\alpha}_b\|^2, \quad (13)$$

Usually, the respiration and its up to third-order harmonics are constructed, i.e., $K_b = 3$ in (7).

Minimizing (14) with respect to the amplitude $\boldsymbol{\alpha}_b$ gives amplitude estimates

$$\hat{\boldsymbol{\alpha}}_b = (\mathbf{H}_b^T \mathbf{H}_b)^{-1} \mathbf{H}_b^T \theta, \quad (14)$$

Inserting $\hat{\boldsymbol{\alpha}}_b$ in (13) yields the estimate of $\hat{f}_b$. Then the estimated $\hat{\mathbf{s}}_{ref} = \hat{\mathbf{H}}_b \hat{\boldsymbol{\alpha}}_b$ is constructed.

Substituting $\hat{\mathbf{s}}_{ref}$ into (12), the subspace matrix $\mathbf{X}$ of respiration and its harmonics is obtained. Minimizing (11) with respect to weight $\mathbf{w}$ gives weight estimates.

$$\mathbf{w} = (\mathbf{X}^T \mathbf{X})^{-1} \mathbf{X}^T \theta, \quad (15)$$

Therefore, the vital sign signal after cancellation becomes

$$\theta_{ECA} = \theta - \mathbf{Xw} = \theta - \mathbf{X}(\mathbf{X}^T\mathbf{X})^{-1}\mathbf{X}^T\theta = \mathbf{P}\theta. \quad (16)$$

where the projection matrix $\mathbf{P}$ projects the received vector $\theta$ in the subspace orthogonal to the respiration subspace and its harmonic subspace.

### C. Adaptive Harmonic Enhanced Trace

The AHET algorithm is proposed as outlined in Algorithm 1 to mitigate intermodulation interference and enhance HR estimation. The credibility assessment criterion is based on the difference between the estimated fundamental HR and its second harmonic frequencies. Specifically, we define the following variables: $\Delta_{f_i}(n)$ as the absolute difference between twice the estimated fundamental HR frequency $f_{h_i}(n)$ and the estimated second harmonic frequency $f_{h_i}^*(n)$; $V_e$ (set to 0.1 Hz) is the deviation threshold that determines the reliability of the frequency estimates; $V_a$ (set to 0.1 Hz) represents the threshold for the difference between two consecutive HR estimates [6]; and $h$ denotes the average of the first five stable HR values. The credibility assessment can be expressed as:

$$\Delta_{f_i}(n) = |2 \cdot f_{h_i}(n) - f_{h_i}^*(n)|, \quad (i=1,2) \quad (17)$$

$$\begin{cases} \Delta_{f_i}(n) \leq V_e, & \text{reliable} \\ \Delta_{f_i}(n) > V_e, & \text{unreliable} \end{cases}$$

Within the same processing window $n$, we first search for the largest power peak in the fundamental HR range (0.7 to 2 Hz) and record the frequency $f_{h_1}(n)$. To search for the second HR harmonic corresponds to $f_{h_1}(n)$, we adaptively adjust the second HR harmonic search region from $f_{h_1}(n) - V_e$ to 4 Hz. Within this adjusted range, we search for the largest power peak and record the corresponding frequency as $f_{h_1}^*(n)$. Once the corresponding harmonic components are identified, the credibility assessment criterion is applied to evaluate the reliability of HR candidates.

If the harmonic relationship is deemed valid, $f_{h_1}(n)$ is considered a reliable HR candidate. Otherwise, the reliability of $f_{h_1}(n)$ is questioned, prompting a further search for the second-largest power peak frequency in the fundamental HR range, denoted as $f_{h_2}(n)$. After refining the search region for the second harmonic and searching for $f_{h_2}^*(n)$, $\Delta_{f_2}(n)$ is computed. The reliability of this new candidate is then re-evaluated using the same credibility assessment. In cases where both $\Delta_{f_1}(n)$ and $\Delta_{f_2}(n)$ exceed $V_e$, additional refinement is necessary to stabilize the peak selection. This involves computing the average of the first five stable HR values, $\bar{h}$, and updating the search regions to $\bar{h} \pm V_a$ Hz and $2 \cdot (\bar{h} \pm V_a)$ Hz. Additional refinement is applied in cases where either of the above conditions holds or if there are significant fluctuations between consecutive HR estimates, to ensure stability in the peak selection process.

Finally, the final HR estimate $f_h(n)$ is computed by taking a weighted average of these selected frequencies.

## V. EXPERIMENTAL RESULTS

This section illustrates the experimental system, adaptive HR monitoring results of the proposed method and performance comparison of the proposed method with conventional peak-finding and ANLS methods across accuracy, robustness, and computational efficiency.

**Algorithm 1:** Adaptive Harmonic Enhanced Trace

1 **Input:** $\theta_{ECA}(t)$: the signal obtained by ECA.
2 **Set:** $V_e, V_a$.
3 **for** $n = 1$ **to** $N$ **do**
4  Find $f_{h_1}(n)$ from the fundamental HR range. Refine the second harmonic search region to $[f_{h_1} - V_e, 4]$ Hz, find $f_{h_1}^*(n)$, and compute $\Delta_{f_1}(n)$.
5  **if** $\Delta_{f_1}(n) \leq V_e$ **then**
6   **Output:** $f_h(n) = \frac{1}{2} \cdot f_{h_1}(n) + \frac{f_{h_1}^*(n)}{2}$
7   Apply additional refinement if necessary.
8  **end if**
9  **else**
10   Find $f_{h_2}(n)$, refine the second harmonic search region, find $f_{h_2}^*(n)$, and compute $\Delta_{f_2}(n)$.
11   **if** $\Delta_{f_2}(n) \leq V_e$ **then**
12    **Output:** $f_h(n) = \frac{1}{2} \cdot f_{h_2}(n) + \frac{f_{h_2}^*(n)}{2}$
13    Apply additional refinement if necessary.
14   **end if**
15   **else**
16    **if** $\Delta_{f_1}(n) > V_e \wedge \Delta_{f_2}(n) > V_e$ **then**
17     Compute $\bar{h}$ from the first five stable HR values. Update the search regions to $[\bar{h} \pm V_a]$ Hz and $[2 \cdot (\bar{h} \pm V_a)]$ Hz. Find $f_h'(n)$ and $f_h''(n)$.
18     **Output:** $f_h(n) = \frac{1}{2} \cdot f_h'(n) + \frac{f_h''(n)}{2}$.
19    **end if**
20   **end if**
21  **end if**
22 **end for**

### A. Experimental System

The experiments were carried out using a 77-GHz FMCW radar evaluation board (TI AWR2243Boost) [25] and a real-time data capture module (TI DCA1000EVM) [26]. Additional details about the radar parameters are summarized in Table I. Under these settings, the radar achieves a 100 Hz frame rate and a total bandwidth of 3.99 GHz. With a phase noise of -94 dBc/Hz, the radar system offers high phase sensitivity to detect subtle skin displacements, supported by a robust transmission power of 13 dBm. The measurements were conducted to validate the comparison between the photoplethysmography (PPG) signal measured by oximeter and the radar HB signal. Participants aged 20 to 30 years were recruited for the experiment and sit in front of the radar and breathe normally as shown in Fig. 3.

### B. Adaptive HR Monitoring

Fig.4 shows three typical experimental cases, in which time-varying spectral powers of HR, $3^{rd}$RR (blue line) and intermodulation interferences $f_1$ (yellow), $f_2$ (purple), $f_3$

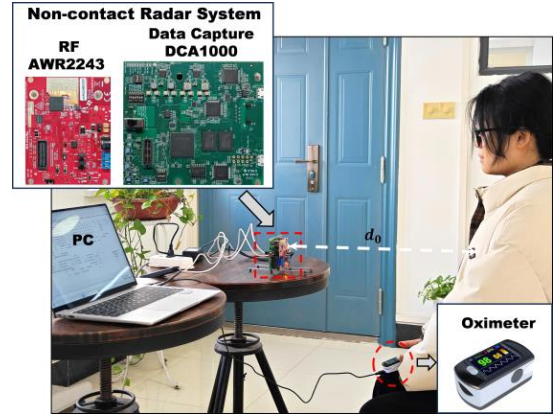

Fig. 3. Example of the experimental scene and data acquisition setup.

(green) in fundamental search range 0.7 to 2 Hz are analyzed by experimental data. Fig.4(a) is the common considered case and only $3^{rd}$RR is comparable to $1^{st}$HR and $1^{st}$HR can be detected after ECA cancelation. The SNR of HR is lower in Fig.4(b), in which both $3^{rd}$RR and $f_1$ are stronger than HR and HR cannot be detected after ECA cancelation. The SNR of HR is the lowest in Fig.4(c), in which $3^{rd}$RR and intermodulation interference $f_1$, $f_2$, $f_3$. How to improve the estimation under these low SNR scenarios is the main scope of this paper.

TABLE I
RADAR PARAMETERS

| Radar Parameters | Value |
|---|---|
| Start Frequency, $f_c$ (GHz) | 77 |
| Frequency Slope, $S$ (MHz/us) | 70 |
| Idle Time (us) | 7 |
| Chirp Time (us) | 50 |
| ADC Samples | 200 |
| ADC Sample Rate (ksps) | 4000 |
| Frame Periodicity, $T_F$ (ms) | 10 |

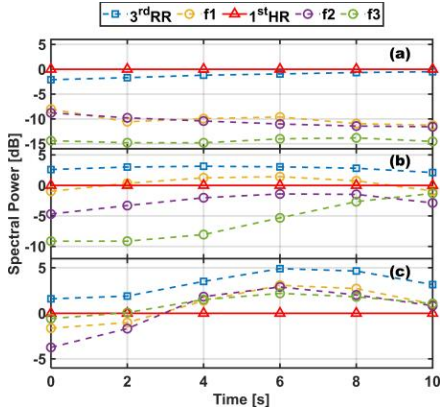

Fig. 4. Time-varying spectral powers of HR, $3^{rd}$RR and intermodulation interference $f_1, f_2, f_3$ in fundamental search range 0.7 to 2 Hz depict three typical experimental cases: (a) Only $3^{rd}$RR is comparable to $1^{st}$HR and the $1^{st}$HR can be detected after ECA cancellation; (b) Both $3^{rd}$RR and $f_1$ are stronger than HR and HR cannot be detected after ECA cancellation; (c) The strength of $1^{st}$HR is lower than $3^{rd}$RR and intermodulation interference $f_1, f_2, f_3$.

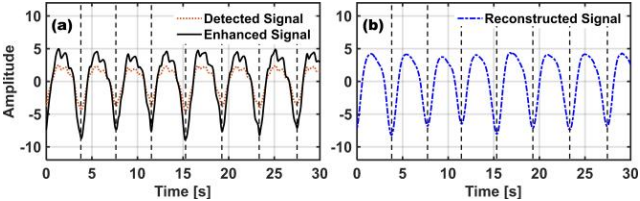

Fig. 5. (a) Original phase-demodulated signal (at the detected range bin) and enhanced signal (from multiple range bins); (b) ANLS-based reconstructed signal, which shows a similar respiratory cycle pattern with enhanced signal.

*1) Respiration and Harmonic Modeling:* Fig.5 shows the respiration and harmonic modeling using ANLS-based signal reconstruction. A sliding-window approach, with a window length of 5 s and a step size of 1 s, is employed as part of the ANLS algorithm for segmental frequency estimation. The ANLS-based reconstructed signal in Fig.5 (b) shows a similar respiratory cycle pattern with enhanced signal in Fig.5 (a).

*2) Respiration Cancellation Results:* Fig.6 shows sliced spectrogram comparisons between before ECA filtered and after ECA filtered corresponding to the three typical experimental cases in Fig.4. It is obvious that HR can be identified in Fig.6(a) after the ECA cancellation. In Fig.6 (b), HR is detectable by checking the visible $2^{nd}$HR. In Fig.6 (c), both the $1^{st}$HR and $2^{nd}$HR are masked by intermodulation interferences. These observations show agreement corresponding to Fig.4. The proposed AHET method is needed to achieve accurate estimation in Fig.6 (b) and (c), which is discussed in the following results.

*3) Adaptive Harmonic Enhanced Trace:* Fig.7 shows HR trace estimation results of a 280s data segment in white dotted line by time-frequency analysis waterfall plots. Fig.7(a) is before filtered and the HR is masked by the $3^{rd}$RR as indicated by the dotted circles. There exists obvious outliers around 120s, 210s and from 250s to 280s. Fig.7(b) is after ECA filtered and the respiration and its harmonics are removed. The detection probability of the fundamental HR has been improved, with the RMSE decreasing from the original 14.41 bpm to 6.37 bpm. However, outliers around 260s still exist due to the intermodulation interference. Fig.7(c) is our proposed AHET result with rmse 1.20 bpm, which shows highly consistent with the PPG reference in Fig.7(d).

## C. Performance Comparison

This subsection aims to evaluate and compare the performance of the proposed HR estimation method with conventional peak-finding and ANLS methods including HR trace estimation, RMSE distributions across various CPIs, and computational time.

Fig.8 illustrates a comparison of HR trace estimation methods over a 280-second measurement period, showing their performance relative to reference values obtained from a PPG device. The methods include the conventional peak-finding approach (blue dashed line), the ANLS method from [16] (green dotted line), and the proposed method (red solid line). The proposed method closely aligns with the reference values, demonstrating superior accuracy and stability throughout the measurement. In contrast, the conventional and ANLS methods exhibit notable deviations, with the errors in the ANLS method primarily due to the initial state of the Kalman filter and the errors between different observations and the true values. This visual comparison highlights the improved accuracy of the proposed method in HR trace estimation.

Table II presents a numerical performance comparison between different methods, with the RMSE (root mean square error) listed for the conventional, ANLS and the proposed AHET methods in different time intervals. The proposed method consistently has the lowest RMSE in most time intervals, indicating better performance.

Fig.9 presents a comparison of RMSE distributions for different methods across multiple CPIs. The methods include conventional (blue violins), ANLS (green violins), and the proposed method (red violins). Each plot corresponds to a specific CPI, illustrating the RMSE variability and central tendency for each method. As the CPI increases, the frequency resolution of the Fourier spectrum improves, thereby reducing the detection errors caused by spectral component aliasing. The proposed method consistently demonstrates the lowest RMSE with minimal variance across all CPIs, indicating superior accuracy and robustness. In contrast, the conventional and ANLS methods show higher RMSE values and greater variability, with the conventional method often having the largest errors. This comparison highlights the effectiveness of the proposed method in reducing estimation errors.

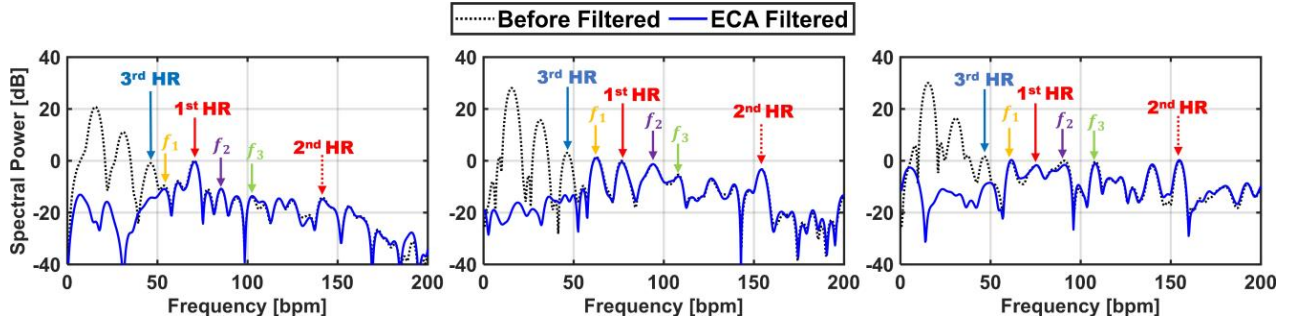

Fig. 6. The sliced spectrogram comparisons between radar detected signal (black dashed line) and after ECA cancellation (blue solid line) and these results are corresponding to the case (a), (b) and (c) in Fig.4, respectively: (a) the HR can be identified as red arrow pointed; (b) the $1^{st}$HR is masked by $f_1$ and the $2^{nd}$HR can be used to justify the detection; (c) $1^{st}$HR is masked by $f_1$ and $f_2$ and $2^{nd}$HR is masked by $f_3$.

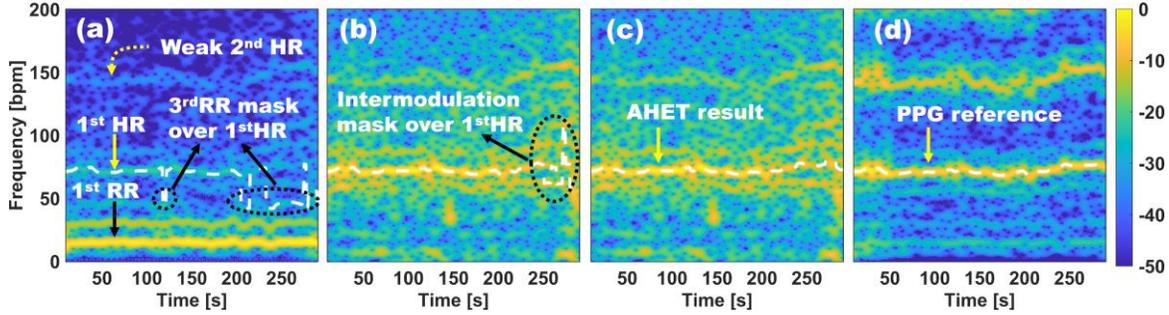

Fig. 7. Time-frequency analysis waterfall plots: (a) is before filtered and the HR is masked by the $3^{rd}$RR as indicated by the dotted circles; (b) is after ECA filtered and outliers still exist due to the intermodulation interference ; (c) is our proposed AHET result; (d) is the PPG reference.

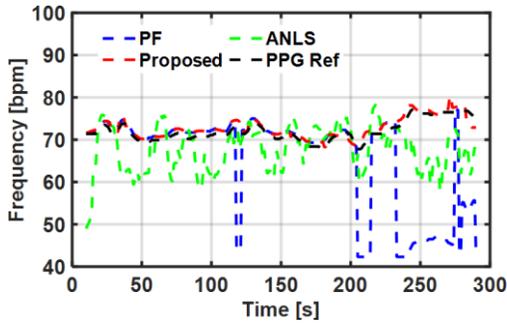

Fig. 8. Different methods of HR trace estimation over a 280-second measurement period using conventional peak finding, ANLS in [16], the proposed AHET method and actual values from the reference PPG device.

TABLE II
PERFORMANCE NUMERICAL COMPARISON BETWEEN DIFFERENT METHODS

| RMSE \ Method Data | Conventional | ANLS | **Proposed** |
|---|---|---|---|
| 10-50s | 0.85 | 10.17 | **0.96** |
| 50-90s | 1.03 | 5.47 | **1.01** |
| 90-130s | 9.01 | 6.82 | **1.21** |
| 130-170s | 0.83 | 5.15 | **0.88** |
| 170-210s | 9.19 | 3.76 | **1.10** |
| 210-250s | 23.39 | 6.17 | **1.53** |
| 250-290s | 27.20 | 11.51 | **1.52** |

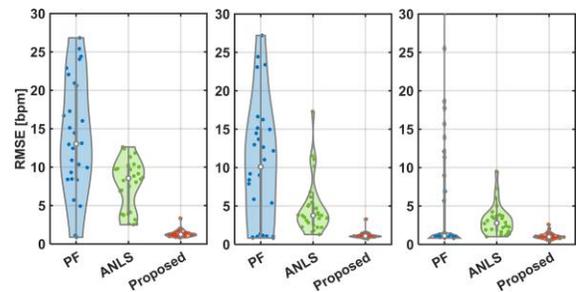

Fig. 9. RMSE comparison of different methods in different CPI: (a) 15s; (b) 20s; (c) 30s.

Fig.10 compares the computational time, normalized to a baseline (conventional peak finding), to process a 280-second of raw radar data. The methods include the baseline conventional peak-finding approach (blue bar), the ANLS method (orange bar), and the proposed method (yellow bar). The conventional method is normalized to 1.000. The ANLS method requires nearly double the time at 1.995, while the proposed method demonstrates improved efficiency over ANLS with a normalized time of 1.305. This comparison highlights that the proposed method achieves a balance between computational efficiency and performance accuracy.

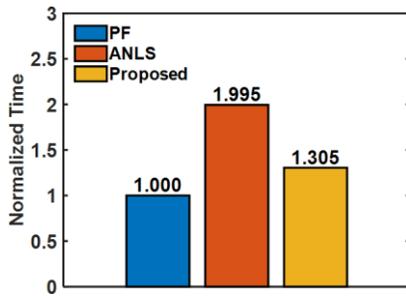

Fig. 10. Computational time of different methods for a single measurement of 280s using conventional peak finding, ANLS in [16], the proposed AHET method.

## VI. Conclusion

This study presents a novel adaptive cancellation before estimation framework in FMCW radar-based HR estimation. To eliminate the respiration and its low-order harmonics, adaptive ECA is successfully implemented with reconstruction by ANLS. To diminish the influence of intermodulation interference and high-order harmonics, AHET method is proposed by defining the credibility assessment between the estimated fundamental HR frequency and the estimated second harmonic frequency to design a fine search region. Experimental evaluations demonstrate superior accuracy, reduced RMSE, and improved computational efficiency. The results highlight the potential of the proposed methods for practical applications in health monitoring, offering a non-invasive, accurate, and robust solution for HR estimation under low SNR conditions.

## References

[1] B. B. Zhang, D. Zhang, Y. Li, Z. Lu, J. Chen, H. Wang, and Y. Chen, "Monitoring long-term cardiac activity with contactless radio frequency signals," *Nature Communications*, vol. 15, no. 1, pp. 1–11, 2024.

[2] Y. Zhang, R. Yang, Y. Yue, E. G. Lim, and Z. Wang, "An overview of algorithms for contactless cardiac feature extraction from radar signals: advances and challenges," *IEEE Transactions on Instrumentation and Measurement*, vol. 72, pp. 1–20, 2023.

[3] A. De Santos Sierra, C. Sanchez Avila, J. Guerra Casanova, and G. Bailador del Pozo, "A stress-eetection system based on physiological signals and fuzzy logic," *IEEE Transactions on Industrial Electronics*, vol. 58, no. 10, pp. 4857–4865, 2011.

[4] M. Mercuri, I. R. Lorato, Y.-H. Liu, F. Wieringa, C. V. Hoof, and T. Torfs, "Vital-sign monitoring and spatial tracking of multiple people using a contactless radar-based sensor," *Nature Electronics*, vol. 2, pp. 252–262, 2019.

[5] P. P. San, S. H. Ling, and H. T. Nguyen, "Industrial application of evolvable block-based neural network to hypoglycemia monitoring system," *IEEE Transactions on Industrial Electronics*, vol. 60, no. 12, pp. 5892–5901, 2013.

[6] Y. Li, C. Gu, and J. Mao, "A robust and accurate FMCW MIMO radar vital sign monitoring framework with 4-D cardiac beamformer and heart-rate trace carving technique," *IEEE Transactions on Microwave Theory and Techniques*, vol. 72, no. 10, pp. 6094–6106, 2024.

[7] C. Li and J. Lin, "Complex signal demodulation and random body movement cancellation techniques for non-contact vital sign detection," in *IEEE MTT-S International Microwave Symposium Digest*, 2008, pp. 567–570.

[8] A. Singh, S. U. Rehman, S. Yongchareon, and P. H. J. Chong, "Multi-resident non-contact vital sign monitoring using radar: A review," *IEEE Sensors Journal*, vol. 21, no. 4, pp. 4061–4084, 2021.

[9] B. Li, W. Li, Y. He, W. Zhang, and H. Fu, "RadarNet: Noncontact ECG signal measurement based on FMCW radar," *IEEE Transactions on Instrumentation and Measurement*, vol. 73, pp. 1–9, 2024.

[10] H. Tang, Y. Rong, L. Chai, and D. W. Bliss, "Deep learning radar for high-fidelity heart sound recovery in real-world scenarios," *IEEE Sensors Journal*, vol. 23, no. 15, pp. 17 803–17 814, 2023.

[11] S. Yao, J. Cong, D. Li, and Z. Deng, "Non-contact vital sign monitoring with FMCW radar via maximum likelihood estimation," *IEEE Internet of Things Journal*, vol. 11, no. 23, pp. 38 686–38 703, 2024.

[12] C. Ma, Q. Shi, B. Hua, Y. Zhang, Z. Xu, L. Chu, R. Braun, and J. Shi, "Noncontact heartbeat and respiratory signal separation using a sub 6 GHz SDR micro-doppler radar," *IEEE Journal of Electromagnetics, RF and Microwaves in Medicine and Biology*, vol. 8, no. 2, pp. 122–134, 2024.

[13] M. Arsalan, A. Santra, and C. Will, "Improved contactless heartbeat estimation in FMCW radar via kalman filter tracking," *IEEE Sensors Letters*, vol. 4, no. 5, pp. 1–4, 2020.

[14] Y. Rong, I. Lenz, and D. W. Bliss, "Noncontact cardiac parameters estimation using radar acoustics for healthcare IoT," *IEEE Internet of Things Journal*, vol. 11, no. 5, pp. 7630 – 7639, 2024.

[15] C. Li, Y. Xiao, and J. Lin, "Experiment and spectral analysis of a low-power $k\alpha$-band heartbeat detector measuring from four sides of a human body," *IEEE Transactions on Microwave Theory and Techniques*, vol. 54, no. 12, pp. 4464–4471, 2006.

[16] G. Beltrão, W. A. Martins, B. Shankar M. R., M. Alaee-Kerahroodi, U. Schroeder, and D. Tatarinov, "Adaptive nonlinear least squares framework for contactless vital sign monitoring," *IEEE Transactions on Microwave Theory and Techniques*, vol. 71, no. 4, pp. 1696–1710, 2023.

[17] G. Beltrão, M. Alaee-Kerahroodi, U. Schroeder, D. Tatarinov, and M. R. Bhavani Shankar, "Statistical performance analysis of radar-based vital-sign processing techniques," in *Sensing Technology*, N. K. Suryadevara, B. George, K. P. Jayasundera, J. K. Roy, and S. C. Mukhopadhyay, Eds. Cham: Springer International Publishing, 2022, pp. 101–112.

[18] B.-K. Park, O. Boric-Lubecke, and V. M. Lubecke, "Arctangent demodulation with dc offset compensation in quadrature doppler radar receiver systems," *IEEE Transactions on Microwave Theory and Techniques*, vol. 55, no. 5, pp. 1073–1079, 2007.

[19] Y. Rong and D. W. Bliss, "Remote sensing for vital information based on spectral-domain harmonic signatures," *IEEE Transactions on Aerospace and Electronic Systems*, vol. 55, no. 6, pp. 3454–3465, 2019.

[20] P. Wang, X. Ma, R. Zheng, L. Chen, X. Zhang, D. Zeghlache, and D. Zhang, "SlpRoF: Improving the temporal coverage and robustness of rf-based vital sign monitoring during sleep," *IEEE Transactions on Mobile Computing*, vol. 23, no. 7, pp. 7848–7864, 2024.

[21] L. Qu, C. Liu, T. Yang, and Y. Sun, "Vital sign detection of fmcw radar based on improved adaptive parameter variational mode decomposition," *IEEE Sensors Journal*, vol. 23, no. 20, pp. 25 048–25 060, 2023.

[22] J. Li, S. Guo, G. Cui, X. Zhou, L. Shi, L. Kong, and X. Yang, "Multidomain separation for human vital signs detection with FMCW radar in interference environment," *IEEE Transactions on Microwave Theory and Techniques*, vol. 72, no. 7, pp. 4278–4293, 2024.

[23] F. Colone, D. W. O'Hagan, P. Lombardo, and C. J. Baker, "A multistage processing algorithm for disturbance removal and target detection in passive bistatic radar," *IEEE Transactions on Aerospace and Electronic Systems*, vol. 45, no. 2, pp. 698–722, 2009.

[24] G. Beltrão, M. Alaee-Kerahroodi, U. Schroeder, D. Tatarinov, and B. S. M. R., "Nonlinear least squares estimation for breathing monitoring using FMCW radars," in *2021 18th European Radar Conference (EuRAD)*, 2022, pp. 241–244.

[25] Awr2243boost. Available: https://www.ti.com/product/AWR2243. Accessed: Jan. 6, 2025.

[26] Dca1000evm. Available: https://www.ti.com/tool/DCA1000EVM. Accessed: Jan. 6, 2025.